\documentclass[pdflatex,sn-mathphys-num]{sn-jnl}

\usepackage{graphicx}%
\usepackage{multirow}%
\usepackage{amsmath,amssymb,amsfonts}%
\usepackage{amsthm}%
\usepackage{mathrsfs}%
\usepackage[title]{appendix}%
\usepackage{xcolor}%
\usepackage{textcomp}%
\usepackage{manyfoot}%
\usepackage{booktabs}%
\usepackage{siunitx}%
\usepackage{algorithm}%
\usepackage{algorithmicx}%
\usepackage{algpseudocode}%
\usepackage{listings}%
\usepackage{lineno}

\usepackage{wrapfig}

\theoremstyle{thmstyleone}%
\theoremstyle{thmstyletwo}%

\theoremstyle{thmstylethree}%

\unnumbered

\begin{document}

\title[Article Title]{Characterization and Active Control of Position-Dependent Timing Dynamics in Superconducting Strip Single-Photon Detectors}

\author*[1,2]{\fnm{Sahil R.} \sur{Patel}}\email{srpatel@jpl.nasa.gov}

\author[3]{\fnm{Kristen M.} \sur{Parzuchowski}}

\author[3]{\fnm{Eli} \sur{Mueller}}

\author[4]{\fnm{Boris} \sur{Korzh}}

\author[1]{\fnm{Emanuel} \sur{Knehr}}

\author[3]{\fnm{Adam N.} \sur{McCaughan}}

\author[3]{\fnm{Martin J.} \sur{Stevens}}

\author[1]{\fnm{Matthew D.} \sur{Shaw}}

\author[1]{\fnm{Jason P.} \sur{Allmaras}}\email{jason.p.allmaras@jpl.nasa.gov}










\affil*[1]{\orgdiv{NASA Jet Propulsion Laboratory}, \orgname{California Institute of Technology}, \orgaddress{\street{4800 Oak Grove Drive}, \city{Pasadena}, \postcode{91109}, \state{CA}, \country{USA}}}

\affil[2]{\orgdiv{Department of Applied Physics and Material Science}, \orgname{California Institute of Technology}, \orgaddress{\street{1200 E. California Blvd.}, \city{Pasadena}, \postcode{91125}, \state{California}, \country{USA}}}

\affil[3]{\orgname{National Institute of Standards and Technology}, \orgaddress{\street{325 Broadway}, \city{Boulder}, \postcode{80305}, \state{CO}, \country{USA}}}

\affil[4]{\orgdiv{Department of Applied Physics Quantum Technologies}, \orgname{Université de Genève}, \orgaddress{\street{Rue de l'Ecole-De-Médecine}, \city{Genève}, \postcode{201205}, \country{Switzerland}}}

\abstract{Superconducting strip single-photon detectors (SSPDs) have emerged as scalable, wide-strip variants of traditional nanowire counterparts. Despite practical advantages—including improved optical fill factors and enhanced signal-to-noise ratios—the fundamental detection physics governing these micro-scale geometries remains largely unexplored. Here, we investigate the underlying photoresponse of a \SI{20}{\micro\meter}-wide tungsten silicide SSPD, demonstrating a slew-rate-corrected timing jitter of 13.2~ps at 532~nm and 20.4~ps at 1550~nm, alongside saturated internal detection efficiency up to 1550~nm. Using focused free-space optical scanning, we reveal that detector timing jitter is strongly influenced by a spatially dependent slew rate between edge and center absorption events. To mitigate this impact, we utilize a parallel superconducting rail architecture to actively redistribute supercurrent. This in-situ tuning minimizes the latency mismatch and mitigates thermally activated intrinsic dark counts, extending the device's ability to  operate at higher temperatures. Finally, comparing these dynamics with time-dependent Ginzburg-Landau (TDGL) modeling elucidates the physical origins of the position-dependent photoresponse, highlighting how superconducting rails or specialized readout electronics can mitigate negative impacts on timing jitter.}

\keywords{SNSPD, SMSPD, SSPD, Rectification, Picosecond Jitter, Tunable Superconductivity}

\maketitle

Superconducting nanowire single-photon detectors (SNSPDs) are a critical technology for advancing quantum information processing \cite{You2020}, scaling photonic quantum computation \cite{Zhong2020}, enabling deep-space optical links \cite{Wollman2024, Hao2024}, and enhancing photon-starved sensing capabilities \cite{Hadfield2023}. The widespread adoption of these detectors is driven by their near-unity system detection efficiency \cite{Reddy2020, Chang2021}, sub-100~ps timing resolution \cite{Korzh2020}, near-zero dark-count rates \cite{Mueller2021}, and high dynamic range \cite{Luskin2023, Allmaras2017}. Traditionally, to achieve practical active areas for these applications the detector is composed of meandered nanowires with widths on the order of tens to hundreds of nanometers. 

Recently, this design paradigm has expanded with the emergence of superconducting strip single-photon detectors (SSPDs)—wide-strip variants of their nanowire counterparts. Early demonstrations of micrometer-scale photon-sensitive bridges \cite{Korneeva2018}, enabled by a greater understanding of the SNSPD detection mechanism \cite{vodolazov2017single}, have since been scaled to active widths of \SI{0.1}{\milli\meter} \cite{Parzuchowski2026}. SSPDs offer practical advantages over nanoscale SNSPD geometries: they improve optical fill factors, enable polarization insensitivity, and facilitate efficient free-space coupling by reducing the insertion losses associated with single-mode optical fibers due to their larger active areas \cite{Parzuchowski2026}. Furthermore, these wide geometries can carry larger bias currents \cite{Chiles2020}. This enhanced current capacity improves the signal-to-noise ratio (SNR)—reducing readout complexity by removing the need for cryogenic amplifiers—and offers the potential for integration with cryo-CMOS and superconducting logic. The ability to efficiently cover large active areas with single, wide strips is advantageous for applications such as direct dark matter detection \cite{Chiles2022, Hochberg2022}, photon-counting LiDAR \cite{McCarthy2025, kuznesof2025midinfrared}, and deep-tissue biomedical imaging \cite{Wang2022,Hughes2026}, where maximizing the collection area is paramount.

However, as the active width scales, the bias supercurrent distribution across the strip becomes non-uniform \cite{Renema2015}. Fabricated devices inevitably possess lithographic edge defects that cause localized geometrical current crowding \cite{Clem2011}. Furthermore, theoretical models predict that the Meissner effect induces additional current crowding at the edges, which becomes increasingly pronounced as the strip width increases \cite{Parzuchowski2026, Gurevich2026, Skocpol1976}. This combination of geometry- and Meissner-induced current crowding restricts the operational switching current and lowers the energy barrier for vortex entry at the edges, facilitating premature, thermally activated intrinsic dark counts \cite{Parzuchowski2026, Gurevich2026}. 

This non-uniform current landscape also affects the detector's photoresponse. Because the local bias current density varies across the width of the strip, the detection dynamics become dependent on the transverse absorption coordinate of the incident photon. This spatial discrepancy generates a latency mismatch between edge and center absorption events, broadening the timing jitter of wide SSPDs \cite{Vodolazov2020_uWire}.

To understand and mitigate this phenomenon, we combine a focused scanning optical spot with a parallel superconducting rail architecture \cite{Parzuchowski2026}. By inductively coupling the SSPD with adjacent control wires, we generate a localized magnetic screening field that counters the out-of-plane component of the strip's self-field (magnetic field vectors perpendicular to the surface of the thin film). Using our optical scanning setup, we first demonstrate that the transverse absorption coordinate is directly encoded into the slew rate (slope) of the output pulse. We subsequently apply active supercurrent redistribution via the rails to manipulate the detector's internal dynamics, minimizing the latency mismatch between edge and center detection events and confirming the spatial origins of the temporal broadening.

Beyond temporal correction, manipulating the localized superconducting response enables new thermal operating regimes. By mitigating localized edge-crowding, we demonstrate that the detector can operate at significantly higher fractions of its critical temperature ($T_c$). This capability not only pushes the thermal limits of conventional amorphous superconductors but also establishes a clear pathway for operating higher-$T_c$ materials, such as NbTiN, at elevated temperatures accessible by compact \SI{4}{\kelvin} cryocoolers.

Ultimately, mapping the intrinsic slew rate to the transverse absorption coordinate establishes a comprehensive physical framework for detection dynamics in micro-scale wide strips. Uncovering this relationship reveals both fundamental consequences and new opportunities for large-area sensing. For photon-number resolution (PNR), this geometric slew-rate variance introduces a masking effect that must be characterized to realize reliable wide-strip PNR architectures. Conversely, it enables the potential for single-strip 2D spatial mapping when combined with established longitudinal time-of-flight techniques \cite{Zhao2017}. Furthermore, mitigating these spatiotemporal constraints via active current redistribution facilitates the deployment of ultra-large-area detectors without sacrificing temporal resolution. To demonstrate these dynamics, we begin by characterizing the spatially resolved photoresponse of a \SI{20}{\micro\meter}-wide WSi device.

\section{Device Characterization and Slew Rate Effects}\label{sec2}

To investigate the underlying photoresponse of SSPDs at the microscale, we characterize a \SI{20}{\micro\meter}-wide and \SI{300}{\micro\meter}-long WSi SSPD \cite{Parzuchowski2026}. Comprehensive details regarding the cryogenic environment, readout electronics, and optical coupling are presented in the Methods section.

\begin{figure}[!ht]
    \centering
    \includegraphics[width=\linewidth]{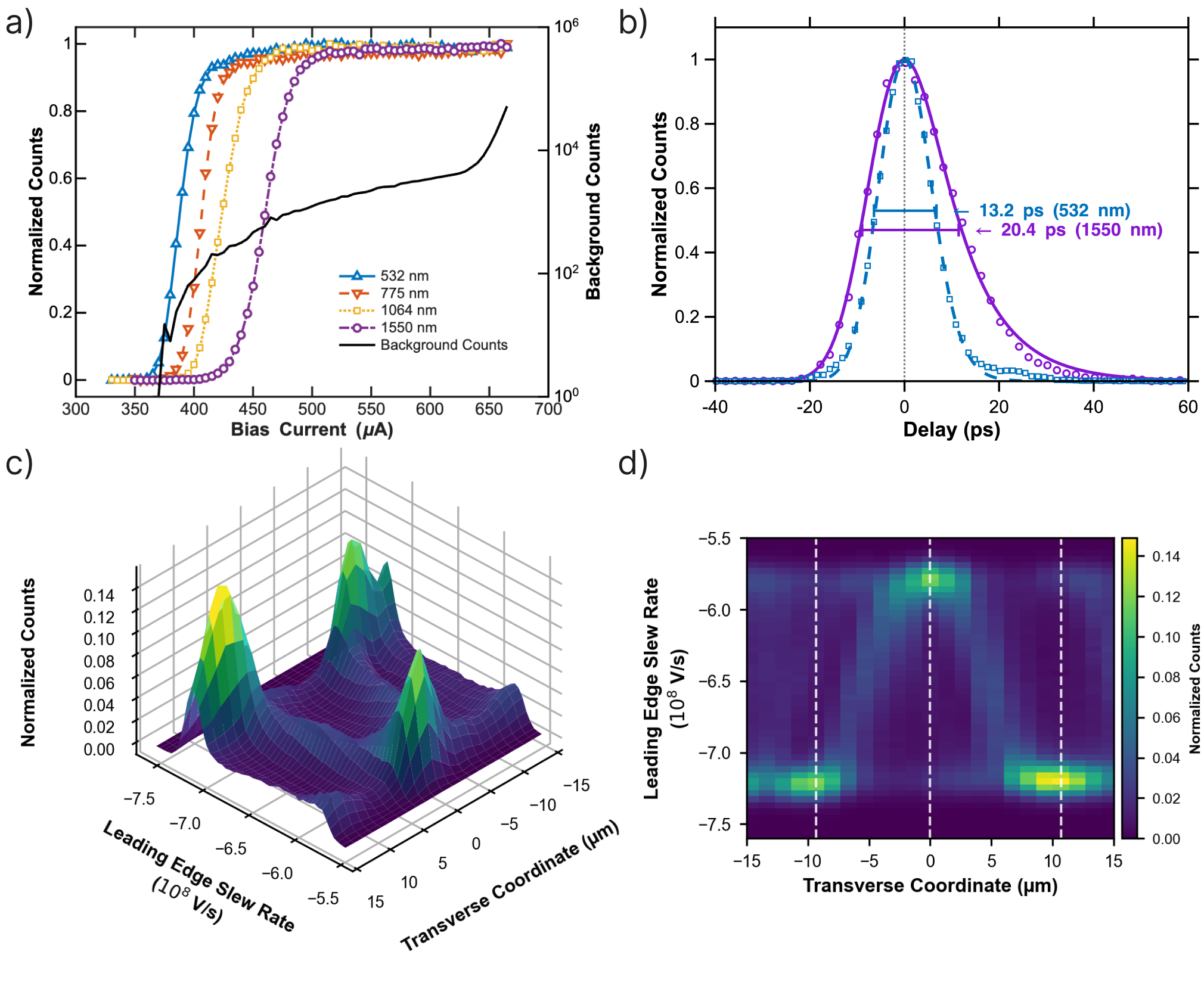}
    \caption{\textbf{Photoresponse and spatial jitter characterization of the\SI{20}{\micro\meter} SSPD.} \textbf({(a)} Photoresponse count rate (PCR) curves taken at four photon wavelengths using a free-space optical setup, demonstrating saturated internal detection efficiency (IDE) up to 1550~nm as indicated by the plateau in photon counts across all wavelengths. Background count rate (BCR) is shown in black corresponding to the right axis. \textbf{(b)} Jitter histograms depicting detector latency at 532~nm and 1550~nm taken at \SI{645}{\micro\ampere}. The blue dashed line and purple solid line represent exponentially modified Gaussian fits to the data. Despite its increased width and use of only room-temperature amplifiers, the slew-rate-corrected jitter remains sub-30~ps, rivaling the performance of much narrower devices. \textbf{(c-d)} Demonstration of location-dependent slew rate in wide SSPD devices. By sweeping a 532~nm wavelength optical spot across the width of the wire, two relatively fast slew rate peaks emerge at \SI{-10}{\micro\meter} and \SI{10}{\micro\meter}, alongside a slower peak at \SI{0}{\micro\meter}, corresponding to edge and center events, respectively.}
    \label{fig:Detector Response}
\end{figure}

The fundamental detector response is shown in Fig.~\ref{fig:Detector Response}a, demonstrating saturated internal detection efficiency (IDE) at wavelengths up to 1550~nm, evidenced by the distinct plateau in the photon count rate. In this saturated regime, every photon absorbed by the superconducting strip triggers a detection pulse. Achieving saturated IDE at these wide transverse length scales overcomes traditional nanoscale geometric constraints.

To probe the detection dynamics at these extended transverse scales, we conducted a spatially resolved latency analysis by systematically translating a focused 532~nm wavelength optical spot across the width of the wire. The beam position was controlled via a room-temperature micrometer that translated rotational increments (degrees) into lateral displacement, resulting in an approximately \SI{20}{\micro\meter} transerse sweep. To resolve the underlying temporal dynamics, we utilized a high-speed (80 Giga samples per second) oscilloscope to capture 32,768 individual pulse traces and their corresponding synchronization signals at each spatial coordinate. By extracting the individual leading-edge slew rates, defined using the 10\%- and 90\%-pulse amplitude crossing points, we compiled the surface map shown in Fig.~\ref{fig:Detector Response}c. In both Fig.~\ref{fig:Detector Response}c and its 2D equivalent Fig.~\ref{fig:Detector Response}d, three distinct groupings emerge, corresponding to detection events at the two edges and the center. The peaks centered at approximately \SI{-10}{\micro\meter} and \SI{10}{\micro\meter} correspond to relatively faster slew rates, indicating edge detection events, whereas the center detections (\SI{0}{\micro\meter}) exhibit noticeably slower slew rates. This observation of a spatially dependent slew rate reveals that the transverse coordinate of photon incidence manifests itself in the shape of the detector output and, as will be seen in later sections, contributes significantly to the temporal resolution of these wide-strip detector architectures.

When correcting for the location-dependent slew rate contribution, the detector timing jitter—defined by the full-width at half-maximum (FWHM) of the latency histogram - is comparable to narrower, traditional SNSPD counterparts. Fig.~\ref{fig:Detector Response}b illustrates the slew-rate-corrected latency histograms, demonstrating 13.2~ps jitter at 532~nm and 20.4~ps jitter at 1550~nm. Achieving this temporal resolution without low-noise cryogenic amplification highlights a practical advantage of scaling to wide SSPD geometries: the inherently large hundreds of microamp operational bias currents yield a high signal-to-noise ratio (SNR), and simplify the required readout architecture.

\section{Consequences of the Transverse Coordinate Effect}\label{sec3}

\begin{figure}[!ht]
    \centering
    \includegraphics[width=\linewidth]{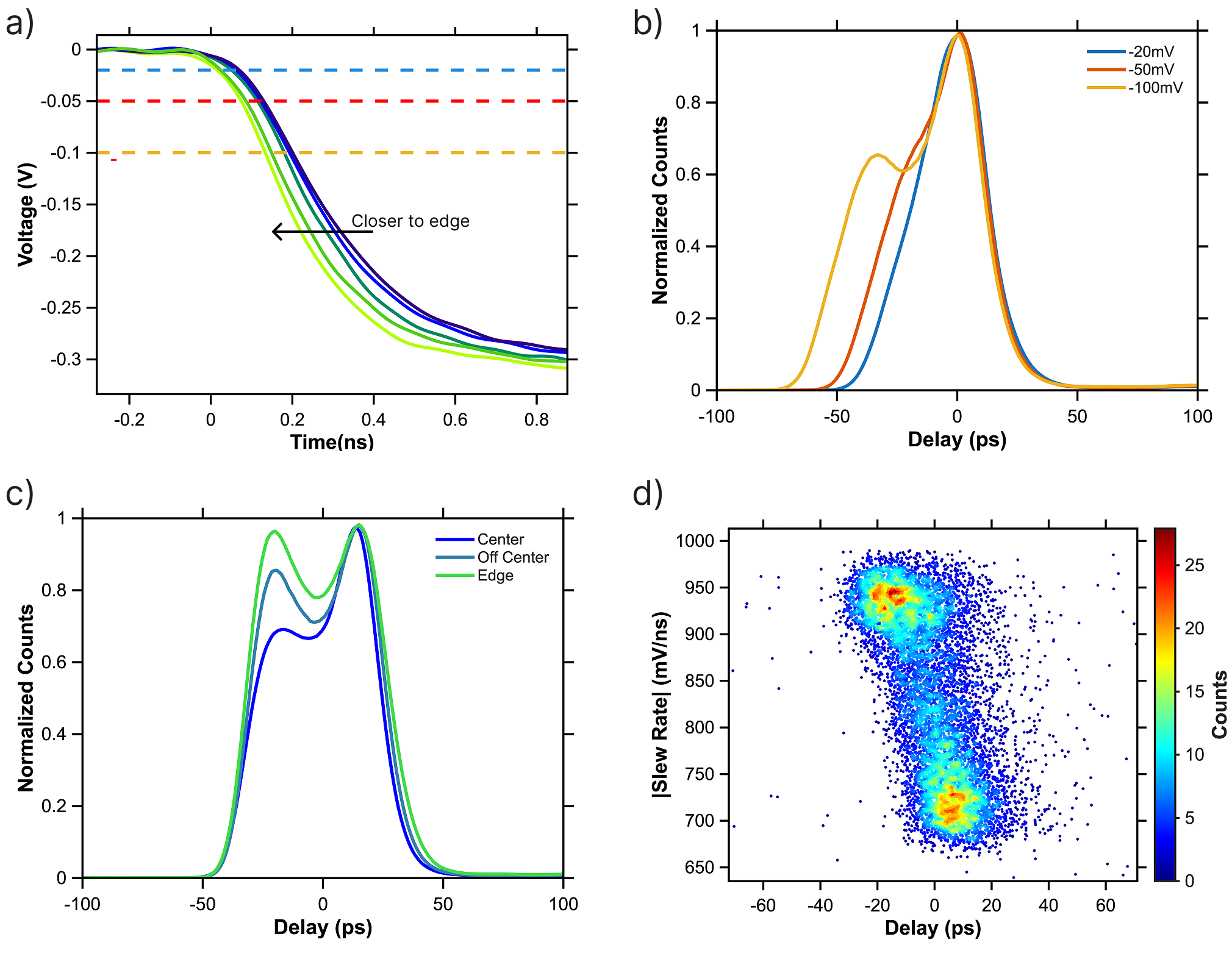}
    \caption{\textbf{Consequences of transverse coordinate dynamics on timing jitter.} \textbf{(a)} Representative voltage pulses taken from the SSPD focused on the leading (falling) edge. Pulses were taken with the optical spot located at the center of the wire. The timing of the signals is with respect to a synchronization signal originating from the pulsed laser source. As the signal progresses further from the onset of the leading edge, the temporal discrepancy between varying slew rates—driven by transverse coordinate effects—becomes more pronounced as shown in panel (\textbf{(b)}). \textbf{(b)} Jitter histograms corresponding to the choice of threshold for illumination at the center of the device. The blue, red, and yellow curves correspond to the \SI{-20}{\milli\volt}, \SI{-50}{\milli\volt}, and \SI{-100}{\milli\volt} thresholds shown as dashed lines in panel (\textbf{a}). As the threshold is set further from the onset of the pulse, there is the emergence of a bimodal distribution and an overall increase in detector jitter. \textbf{(c)} Demonstration of bimodal jitter histogram changes as the photon incidence location is shifted at a fixed threshold of \SI{-100}{\milli\volt}. As the illumination spot is moved closer to the edge of the wire, we observe proportional growth in the peak near \SI{-20}{\pico\second}, which corresponds to edge events. \textbf{(d)} Oscilloscope trace analysis of the pulses taken on the detector using a threshold of \SI{-100}{\milli\volt} and stationary optical spot located near the edge of the wire. Each point in the panel is a single pulse taken from the SSPD and the slew rate is calculated using the 10\% and 90\% points on the slope. As the delay is sensitive to threshold we used a fixed \SI{-100}{\milli\volt} threshold to calculate the delay between the pulse and the sync signal from the laser. Using this trace analysis technique two clear groupings of pulses are visible, corresponding to edge detections at faster slew rates and center detections at slower slew rates. The delay between the two groupings correlates to the timing difference between the two peaks in the bimodal jitter histogram when using thresholds further from the pulse onset (\SI{-100}{\milli\volt} in this instance).}
    \label{fig:Transverse Coordinate Consequence}
\end{figure}

For standard SNSPD operation, a fixed voltage threshold is typically used to register output pulses upon photon detection. In conventional SNSPDs, the choice of a fixed threshold is important for overall jitter properties, yet it does not significantly distort the shape of the jitter histogram. Wide SSPDs, by contrast, exhibit a markedly different behavior. Driven by the spatially dependent slew rate, a bimodal jitter histogram emerges depending on the choice of trigger threshold. Fig.~\ref{fig:Transverse Coordinate Consequence}a illustrates representative pulses demonstrating varying slew rates, which cause the pulses to temporally separate as they progress further from the onset of the leading edge. Correlating with the scanning measurements (Fig.~\ref{fig:Detector Response}c), the faster slew rates indicate edge detection events, while slower slew rates correspond to center events.

This transverse-coordinate-dependent variation in slew rate dictates the shape of the jitter histogram as a function of the chosen threshold level. As depicted in Fig.~\ref{fig:Transverse Coordinate Consequence}b, shifting the threshold further from the signal onset alters the jitter distribution from a nearly Gaussian to a bimodal shape. The two peaks in this bimodal structure correspond to the edge (faster slew) and the center (slower slew) of the device. This spatial correlation is explicitly verified in Fig.~\ref{fig:Transverse Coordinate Consequence}c; translating the illumination spot toward the strip edge causes proportional growth in the counts associated with the earlier edge peak. Furthermore, voltage trace analysis (Fig.~\ref{fig:Transverse Coordinate Consequence}d) reveals two distinct clusters in the delay versus slew rate parameter space. The clear separation of these clusters, alongside the delay between them, further corroborates the mechanism behind the bimodal jitter structure.

In practice, if the readout electronics can threshold close to the signal onset, the transverse jitter effects arising from slew rate dependence can be mitigated. However, this phenomenon still poses a limitation for applications requiring high temporal precision, such as QKD and deep-space optical communications. In these fields, minimizing timing jitter is critical for scaling pulse-position modulation to higher data rates \cite{Wollman2024} and extending secure network distances in fiber-based quantum key distribution (QKD) protocols \cite{Chen2021,Zadeh2020,Grunenfelder2023}. Furthermore, while fluctuations in the longitudinal position of photon absorption are known to induce geometric jitter via RF propagation delays, this transverse coordinate effect introduces a distinct, additional spatial contribution \cite{Calandri2016}. By compounding existing electronic and longitudinal geometric contributions, the transverse slew rate variance imposes an additional barrier to achieving the intrinsic temporal resolution limits of wide-strip architectures \cite{allmaras2019intrinsic}. Crucially, this spatial variance in the slew rate poses a fundamental challenge for implementing photon-number-resolving (PNR) architectures using wide-strip geometries. Because temporal PNR readouts typically rely on differentiating the steeper slew rates generated by simultaneous multi-photon absorption events \cite{Cahall2017, Nicolich2019}, a spatially dependent single-photon slew rate originating at the strip edge could easily mimic or mask multi-photon signatures. Furthermore, as PNR architectures scale to highly multiplexed arrays \cite{Cheng2023}, decoupling these transverse delays will be essential for integrating wide-strip geometries into next-generation photon-number-resolving systems.

\section{Utilizing Superconducting Rail to Probe Transverse Coordinate Dynamics}\label{sec4}

One promising approach to engineering the transverse coordinate landscape is the integration of superconducting rails, as proposed by Gurevich et al. \cite{Gurevich2026} and implemented by Parzuchowski et al. \cite{Parzuchowski2026}. The rail consists of a higher-$T_c$, thicker superconducting film running parallel to the SSPD, with the rail current biased in the same direction as the SSPD bias. The rail reduces the self-field density at the edges of the SSPD, pushing the bias current inward and away from edge defects that typically cause premature switching.  When properly tuned, modeling shows this produces a more uniform current profile across the wire \cite{Parzuchowski2026,Gurevich2026}. Beyond improving the switching current, we demonstrate that the rail can be utilized to actively tune the timing dynamics of the SSPD, effectively counteracting the location-dependent slew rate and suppressing the transverse coordinate effect in wide-strip detectors.

\begin{figure}[!ht]
    \centering
    \includegraphics[width=\linewidth]{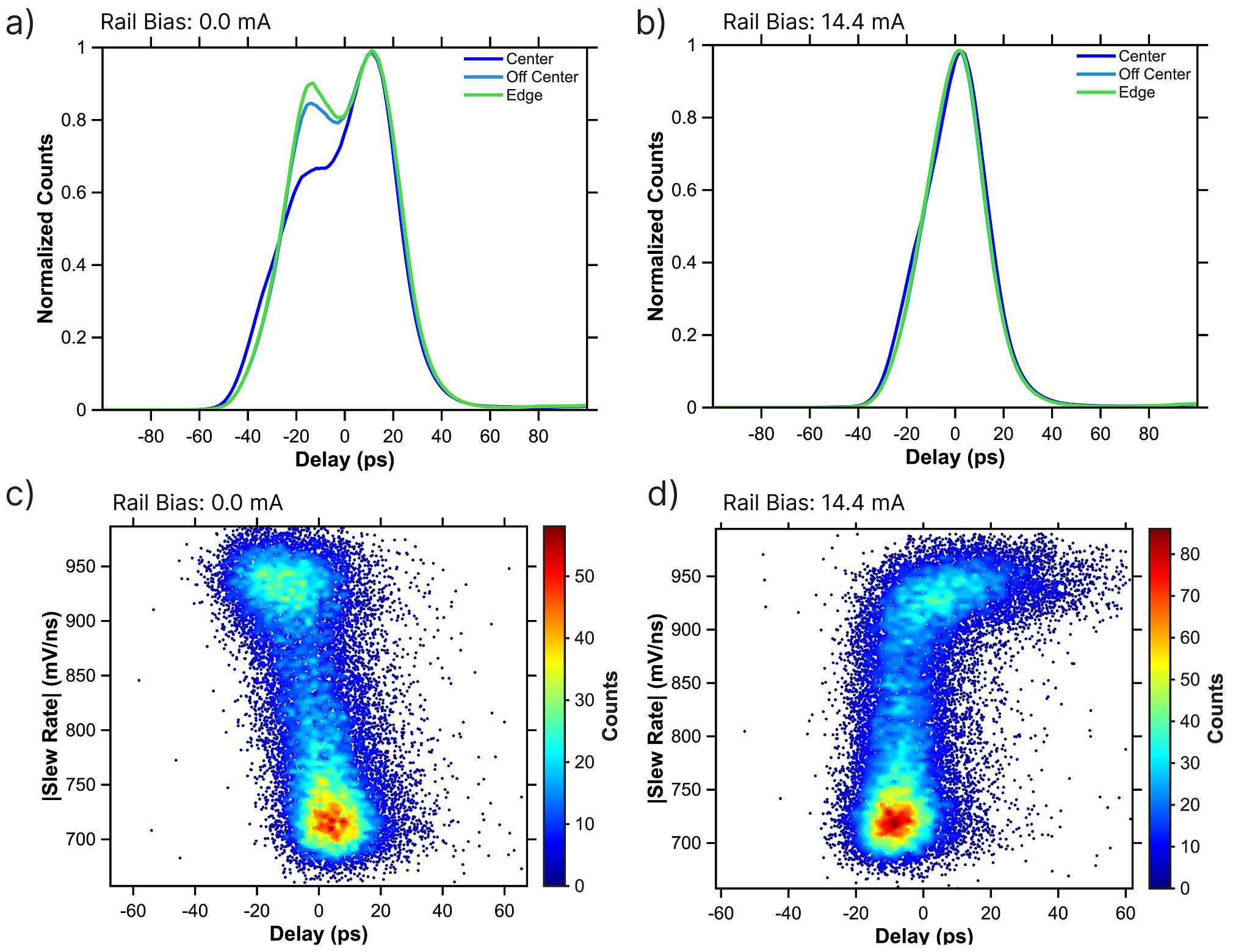}
    \caption{\textbf{Active suppression of position-dependent timing jitter via superconducting rail tuning.} \textbf{(a)} Typical SSPD jitter response at a \SI{-100}{\milli\volt} threshold under a focused optical spot of 1550~nm wavelength with \SI{0}{\milli\ampere} rail current applied. As the optical spot is moved closer to the edge, the secondary faster peak (corresponding to edge-detection events) rises in relative intensity. \textbf{(b)} At a \SI{14.4}{\milli\ampere} rail current, the jitter resolves into a primarily Gaussian distribution leading to a reduction in the FWHM going from 48.8~ps to 28.4~ps. The bimodal structure is suppressed, and moving the spot toward the edge minimizes the induction of a secondary peak. \textbf{(c)} Oscilloscope trace distributions taken at \SI{0}{\milli\ampere} rail current with the spot near the center of the detector. A high density of counts is visible at slower slew rates (center events), separated temporally from the faster edge events. \textbf{(d)} Oscilloscope trace distribution after applying the \SI{14.4}{\milli\ampere} rail current using the same center location and threshold as in (\textbf{c}). The latency mismatch between the two clusters converges. Notably, the higher slew rates corresponding to edge events have a greater distribution of points extending to longer delay values. The shift towards longer delays for higher slew rates and therefore events at the edge correlates with the reduced local current density at the edges induced by the rail's magnetic field.}
    \label{fig:RailJitterIntervention}
\end{figure}

This active timing tuning is observed in the evolution of the detector jitter as a function of rail current, as shown in Figs.~\ref{fig:RailJitterIntervention}a-b. In Fig.~\ref{fig:RailJitterIntervention}a, the detector operates at a fixed bias current of \SI{645}{\micro\ampere}, a typical operating point on the plateau of the 1550~nm photoresponse count rate (PCR) curve, with no rail current applied. As in Fig.~\ref{fig:Transverse Coordinate Consequence}c, using a fixed \SI{-100}{\milli\volt} threshold, the distinctive bimodal jitter structure is evident and as the illumination spot moves towards the edge, the secondary, faster peak increases in relative intensity. In Fig.~\ref{fig:RailJitterIntervention}b, a rail current of \SI{14.4}{\milli\ampere} is applied, and the same spatial scan is performed. Here, the bimodal edge and center peak structure is suppressed, leaving a primarily Gaussian distribution with a reduced full-width at half-maximum (FWHM), thereby minimizing the location-dependent jitter.

To confirm whether the superconducting rail rectification strategy redistributes current away from the edges and mitigates the spatial discrepancy for photon detection, we again utilized oscilloscope trace analysis. Fig.~\ref{fig:RailJitterIntervention}c displays a colormap of the slew rate versus timing delay taken with \SI{0}{\milli\ampere} of rail current, with the optical spot placed centrally (similar to the conditions in Fig.~\ref{fig:RailJitterIntervention}a). The characteristic delay discrepancy between the center and edge detection clusters is clearly visible. However, when the rail current is increased to \SI{14.4}{\milli\ampere} (Fig.~\ref{fig:RailJitterIntervention}d), the delay mismatch between the edge and center slew rates decreases. This mitigation suppresses the bimodal structure and reduces the location dependence. At the faster slew rates corresponding to edge events, there is a measurable increase in the number of events with longer delays. This shift correlates with a reduction in the local bias current at the edges, a direct consequence of the field induced by the rail. The lower bias current density at the edges leading to higher detector delays is well documented in Korzh et al. \cite{Korzh2020,Sidorova2017}. Conversely, when current is supplied to the rail in the opposite direction, the location-dependent slew rate amplifies, and we observe an increased temporal separation in the edge jitter. Ultimately, these investigations highlight the potential of utilizing superconducting rails to actively tune internal detector dynamics and recover optimal timing performance in large-area SSPDs.

\section{Time-Dependent Ginzburg-Landau Model}\label{sec5}

To understand the slew rate dependence on transverse coordinate, we use the combined time-dependent Ginzburg-Landau (TDGL) and two-temperature model \cite{vodolazov2017single, Vodolazov2019, allmaras2019intrinsic, allmaras2020thesis} to qualitatively simulate the superconducting to normal state transition following photon detection. Simulating a full \SI[number-unit-product=\text{-}]{20}{\micro\meter}-wide wire is not practical for computational reasons, but the qualitative features observed experimentally can be observed in the simulated dynamics of a narrower wire.

\begin{figure}[!ht]
    \centering
    \includegraphics[width=\linewidth]{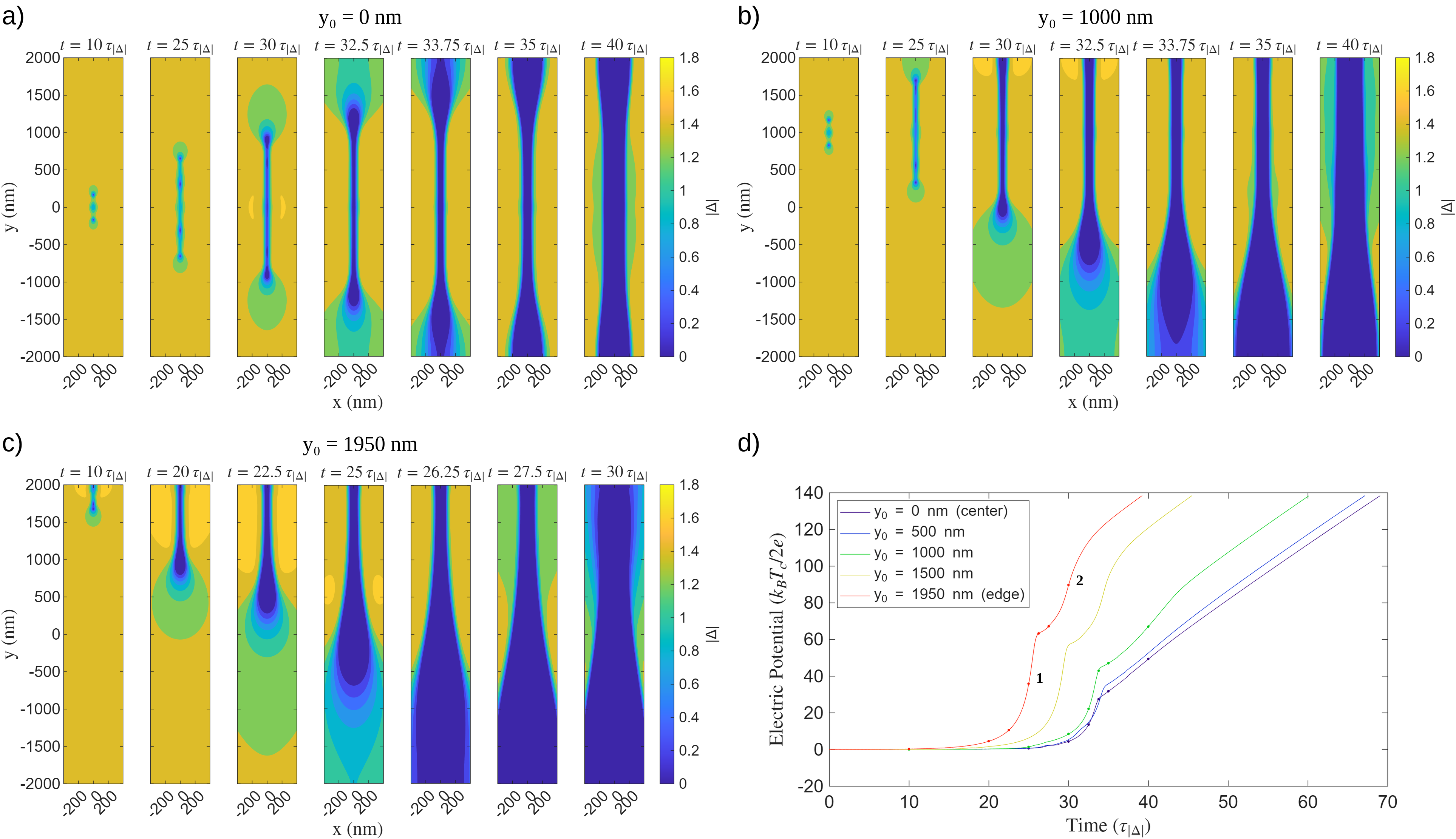}
    \caption{\textbf{Time-dependent Ginzburg-Landau (TDGL) simulation of wide-strip photoresponse.} \textbf{(a–c)} Simulated order parameter magnitude following photon absorption for detection events at (\textbf{a}) center ($y_0 = 0$ nm), (\textbf{b}) intermediate ($y_0 = 1000$ nm), and (\textbf{c}) edge ($y_0 = 1950$ nm) locations for a \SI[number-unit-product=\text{-}]{4}{\micro\meter}-wide WSi wire. \textbf{(d)} Voltage transient response of TDGL model for different hotspot transverse coordinates. The points on the voltage traces correspond to times of the snapshots of the order parameter magnitude shown in (\textbf{a--c}). The `1' and `2' indicators on the $y_0 = 1950$ nm trace highlight the two distinct regimes of increased slew rate as discussed in the text.}
    \label{fig:TDGL}
\end{figure}

Fig.~\ref{fig:TDGL}d shows the simulated voltage transients, with time in units of $\tau_{|\Delta|} = \hbar/k_B T_c$, of a \SI[number-unit-product=\text{-}]{4}{\micro\meter}-wide WSi wire for photon absorptions at different transverse locations $y_0$, with $y_0 = \SI{0}{\nano\meter}$ defined as the center of the wire. The detection event near the edge of the wire features both a shorter latency between absorption and the onset of the voltage transient and a faster initial increase in voltage compared to detection events in the center or intermediate area of the wire. As shown in Figs.~\ref{fig:TDGL}a–c, the breakdown of superconductivity occurs in an expanding front as vortices traverse the width of the wire. Current redistribution and crowding around the leading edge of the normal domain front and the associated dissipation cause this expansion in the lengthwise direction of the wire. As current flows around the leading edge, the local current density exceeds the depairing current density along a wide angular range at the leading edge, not just at the outermost location in the direction of vortex motion. For detection events that begin near the edge of the wire, this expansion takes place over more of the wire's width as the vortex front traverses the wire, leading to the formation of a wider normal domain and a more rapid increase in electric potential, as shown in panels 4 and 5 of Fig.~\ref{fig:TDGL}c compared to the similar panels in Fig.~\ref{fig:TDGL}a. This is the observed larger initial slew rate for detections near the edge of the wire. In contrast, the detections in the center of the wire have the least amount of current-redistribution-associated expansion of the normal region.  The `1' label in Fig.~\ref{fig:TDGL}d highlights the area of the voltage transient that corresponds with this initial normal domain expansion.

There is a secondary effect that contributes to the transverse-coordinate-dependent slew rate in the simulations. Immediately after the normal domain extends across the width of the wire, current redistributes based on the geometry of the newly formed normal domain and concentrates in the areas with the narrowest normal domain along the length of the wire. For the edge detections where the asymmetry in the normal domain shape is largest, this leads to a significant redistribution of current and an increased local current density flowing through the normal region. The local Joule heating, proportional to $j^2 \rho$, is enhanced relative to a uniform current distribution along the width of the wire, leading to a more rapid expansion of the normal domain, and a corresponding increase in slew rate in the electric potential, labeled as `2' in Fig.~\ref{fig:TDGL}d. The voltage trace of the intermediate location of $y_0 = \SI{1000}{\nano\meter}$ shows the same effect following the initial motion of flux across the width of the wire. In the model, the rate of normal domain growth stabilizes to a constant value for all initial coordinates once this non-uniform normal domain growth equalizes the transverse coordinate differences in normal domain length. At that point, the domain grows as a 1D front along the length of the wire. Further details of the simulation are described in Methods:Time-Dependent Ginzburg-Landau Model.

While the \SI[number-unit-product=\text{-}]{4}{\micro\meter}-wide results are not quantitatively representative of the measured \SI[number-unit-product=\text{-}]{20}{\micro\meter}-wide wire, the normal domain formation and current redistribution are anticipated to exhibit the same qualitative behavior. The features of expanding normal domain as the domain traverses the width and current redistribution following the full formation of the normal domain are general features of wide wires. The model also helps explain how the application of superconducting rail current reduces the timing jitter of these devices. By reducing the current density near the edges of the wire, the initial vortex formation and velocity are slower, which delays the initial voltage formation relative to events triggered in the center of the wire. As this initial delay becomes larger than that of events in the center of the wire, the time delay difference from the slew rate difference is partially mitigated.

\section{Higher Temperature Operation}\label{sec6}

To explore the utility of the rail architecture beyond timing correction, we investigated its impact on the elevated-temperature performance of the wide SSPD. As the operating temperature approaches the critical temperature ($T_c$), superconducting detectors suffer performance degradation driven by a decrease in the energy barrier for vortex entry. This leads to an exponential increase in intrinsic material noise (dark count rate).

\begin{figure}[!ht]
    \centering
    \includegraphics[width=\linewidth]{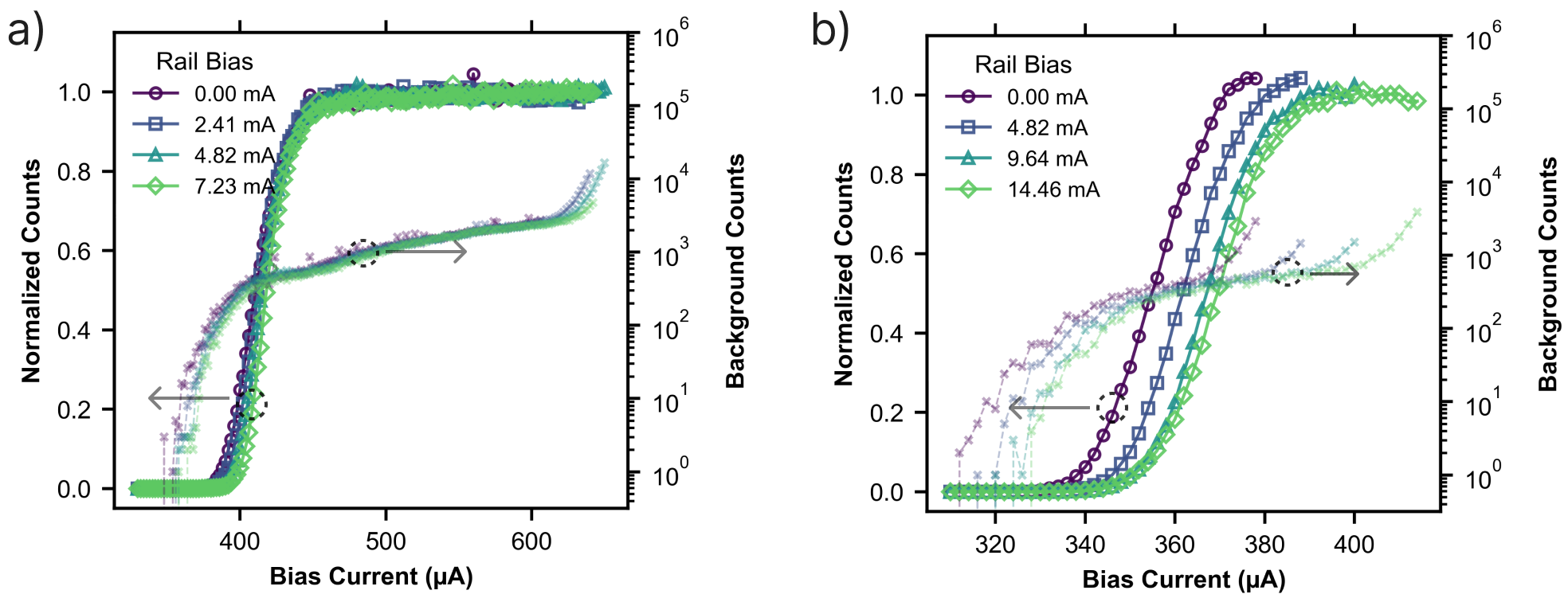}
   \caption{\textbf{Elevated-temperature operation enabled by active current redistribution.} \textbf{(a)} Subtracted photoresponse count rate (PCR) and dark count rate (DCR) at \SI{1.0}{\kelvin} under 1064~nm illumination. The detector demonstrates a distinct plateau across varying active rail currents. \textbf{(b)} Detector response at an elevated temperature of \SI{1.625}{\kelvin}. Without active current redistribution (\SI{0}{\milli\ampere} rail current), thermally activated vortex entry increases intrinsic DCR sharply, collapsing the plateau. Applying a rail current (up to \SI{14.5}{\milli\ampere}) mitigates edge-crowding and systematically shifts the DCR onset, recovering a viable operational plateau.}
    \label{fig:HighTemp}
\end{figure}

At a baseline temperature of \SI{1.0}{\kelvin} ($\sim0.3T_c$) (Fig.~\ref{fig:HighTemp}a), the device exhibits a distinct plateau in the 1064~nm subtracted photon count rate, with intrinsic dark counts remaining negligible until the bias current closely approaches the critical switching current (exponential increase regime). However, when the temperature is elevated to \SI{1.625}{\kelvin} ($\sim0.5T_c$)(Fig.~\ref{fig:HighTemp}b) with no active current redistribution, the intrinsic dark count rate (DCR) increases sharply at lower bias currents. This thermally activated vortex entry at the edges collapses the viable operating plateau \cite{Bulaevskii2011}. 

By applying an active rail current, we mitigate localized current crowding at the edges where these premature vortices preferentially enter, systematically shifting the onset of the DCR to higher bias currents and restoring a viable detection plateau at \SI{1.625}{\kelvin}. The localized heating introduced by the active rail remains well within the thermal budget of the cryogenic stage, ensuring that the active redistribution does not introduce prohibitive parasitic thermal loads. This capability establishes a clear pathway for robust, high-temperature operation of wide-strip architectures without sacrificing the spatiotemporal tuning enabled by the rail.

\section{Conclusion}\label{sec_conclusion}
This work establishes that the temporal resolution of wide-strip WSi SSPDs is strongly influenced by a spatially dependent slew rate. With focused free-space optical scanning, we demonstrated that the transverse coordinate of photon incidence dictates the slew rate of the output pulse. This spatial discrepancy is the primary mechanism driving the emergence of macroscopic bimodal jitter in micrometer-scale geometries. The time-dependent Ginzburg-Landau (TDGL) simulations provide a qualitative framework for this phenomenon, linking the macroscopic voltage response to the underlying vortex dynamics and localized Joule heating during normal domain expansion. Additionally, we demonstrated how a parallel rail architecture provides an active, secondary means to mitigate these transverse effects by manipulating the current density in the superconducting strip.

Understanding this geometric noise source provides immediate insight and utility for the design and readout of SSPDs. Most notably, elucidating the origin of this bimodal jitter provides the community with a mechanistic foundation for optimizing readout thresholding in wide-wire devices. By setting voltage thresholds near the pulse onset or using a two-threshold system to simultaneously extract the slew rate, researchers can minimize the influence of transverse spatial delays. This mitigation is critical for achieving the precise system timing jitter required for large-area, low-flux applications, such as deep-space optical communications and long-range LiDAR.

Beyond temporal optimization, the relaxation of nanoscale geometries allows for the fabrication of large active areas with near-unity optical fill factors. The transition from meandering nanowires to 2D micro-strips inherently enables polarization-insensitive detection, a highly desirable trait for unguided optical communication and astronomy. Furthermore, the large operational bias currents supported by these wide geometries yield a high SNR. This enhanced SNR is exceptionally well-suited to drive advanced multiplexing architectures, such as the thermally coupled imager (TCI) \cite{McCaughan2022_TCI}. Integrating wide-strip SSPDs into a TCI framework—where detector pulses are required to trigger a localized thermal readout bus—could enable scalable, megapixel-resolution spatial imagers utilizing minimal microwave readout lines. Such large-format arrays would provide a critical enabling technology for photon-starved disciplines, including dark matter haloscope searches and astronomy.

Looking forward, the fundamental understanding of spatial-to-temporal mapping established here provides a structural blueprint for the next generation of engineered single-photon detectors. Future investigations could utilize high-resolution optical scanning to precisely map localized slew rate variations arising from variations in current density across even wider architectures or different material systems. Furthermore, integrating this active current redistribution strategy with emerging micro-scale high-$T_c$ superconducting films, such as niobium-titanium nitride \cite{Yabuno2023, Haneishi2025}, presents a grounded pathway to mitigate thermally activated dark counts at elevated temperatures. Ultimately, by transforming fundamental geometric constraints into tunable parameters, this work unlocks the broader potential of wide-strip geometries, paving the way for scalable, ultra-large-area quantum sensors.

\backmatter

\section*{Data availability}
The data that support the plots within this paper and other findings of this study are available from the corresponding author upon reasonable request.

\section*{Code availability}
The time-dependent Ginzburg-Landau (TDGL) simulation code used in this study is not available for public distribution. Inquiries regarding its structure and performance can be directed to the corresponding author.

\section*{Acknowledgments}
Part of this research was performed at the Jet Propulsion Laboratory, California Institute of Technology, under contract with the National Aeronautics and Space Administration (80NM0018D0004). Support for this work was provided by the DARPA SynQuaNon program. NIST work was funded solely by the US government. \textcopyright 2026. All Rights Reserved. Government sponsorship acknowledged.  This document has not been peer reviewed but has been cleared by NIST for release. The use of trade names is intended to allow the
measurements to be appropriately interpreted and does not imply
endorsement by the US government, nor does it imply these are
necessarily the best available for the purpose used here.

\section*{Author contributions}
S.R.P. and J.P.A. conceived the experiment and S.R.P. performed the measurements. J.P.A. developed the TDGL framework and performed the simulations. The Faint Photonics Group at NIST Boulder (A.N.M., E.M., K.M.P., M.J.S.) designed and fabricated the devices. B.K., E.K., J.P.A., and S.R.P. contributed to the cryogenic setup and measurement infrastructure. M.D.S. and M.J.S. provided funding support and project conception. All authors discussed the results and contributed to the writing of the manuscript. We thank Emma Wollman for fruitful discussions regarding the results.

\section*{Competing interests}
The authors declare no competing interests.

\section*{Additional information}
\textbf{Correspondence and requests for materials} should be addressed to Sahil R. Patel.

\begin{appendices}

\section{Methods}
\subsection{Cryogenic Setup and Readout Electronics} 
The SSPD was tested in a He-4 sorption fridge with a base temperature of \SI{0.9}{\kelvin}. The detector was biased and its output pulses were read out utilizing a room-temperature bias-tee and a low-noise amplifier (Mini-Circuits ZFL-1000LN+). Photoresponse count rates (PCR) were recorded using a Swabian Time Tagger Ultra, and raw pulse traces were captured with a high-speed Keysight oscilloscope. To prevent latching—a condition where the device becomes locked in a stable resistive state following photon absorption—an additional \SI{5}{\ohm}, \SI{470}{\nano\henry} inductive shunt was integrated into the readout circuit.

\subsection{Optical Coupling and Spatial Scanning} 
The device was free-space coupled through BK7 windows at the \SI{300}{\kelvin} and \SI{40}{\kelvin} stages, with additional \SI{1.9}{\micro\meter} and \SI{1.6}{\micro\meter} angled short-pass filters positioned at the \SI{4}{\kelvin} and \SI{40}{\kelvin} stages, respectively. As previously demonstrated \cite{Mueller2021}, this specific filtering configuration reduces background blackbody counts (BCR) and facilitates optimal device operation. We utilized a free-space optical coupling methodology \cite{Korzh2020} to illuminate the active area using \SI{1064}{\nano\meter}/\SI{532}{\nano\meter} and \SI{1550}{\nano\meter}/\SI{775}{\nano\meter} frequency-doubled laser sources. During alignment, strict care was taken to center the beam on the active area while avoiding the tapered regions. The detector count rate served as a secondary alignment metric, with the wire edges verified by a $>50\%$ decrease in the count rate ($\sim$10~kcounts/s). Once aligned, the beam was attenuated to the single-photon regime ($\sim$30~kcounts/s) using a combination of neutral density filter wheels and absorptive filters.

\subsection{Time-Dependent Ginzburg-Landau Model}\label{secA_TDGL}

The TDGL simulations were performed based on the formulation of Vodolazov \cite{vodolazov2017single}, with the modifications discussed in \cite{allmaras2019intrinsic} and \cite{allmaras2020thesis}. The model assumes material properties similar to the WSi used in this work. For the TDGL equations, the material parameters are critical temperature $T_c = \SI{3.1}{\kelvin}$, electronic diffusion coefficient $D = \SI{0.7}{\centi\meter^2/\second}$, sheet resistance $\rho_{sq} = \SI{843}{\ohm/sq}$, substrate temperature $T_{sub} = T_c/2$, and $\tau_{ee}(T_c) = \SI{0}{\pico\second}$, which reduces the equations to the standard rather than generalized TDGL.  For the two-temperature energy balance equations, the additional material parameters are electron-phonon coupling $\tau_0 = \SI{13.1}{\nano\second}$, $\tau_{esc} = \SI{300}{\pico\second}$, phonon parameter $\gamma=94$, and down-conversion coupling time constant $\tau_{DC} = \SI{0.62}{\pico\second}$. The grid size is fixed at \SI{4}{\nano\meter}, and the width is \SI{4000}{\nano\meter}.

The bias current ($I_B$) is set to $I_B = 0.8 I_{dep}(T)$ where  $I_{dep}(T)$ is the temperature dependent depairing current. The simulated nanowire is connected to a circuit model consisting of a \SI{1}{\micro\henry} series inductor and \SI{50}{\ohm} shunt resistor.  Given the large series inductance, the nanowire behaves as a nearly fixed-current biased device with little current shunting during the timescale of the simulation. Photon energy is deposited into the electron system using a decaying exponential rate ($\tau_{DC}$) justified by the solution of the kinetic equations for the initial down-conversion cascade \cite{vodolazov2017single} and the phonon plateau initial conditions \cite{allmaras2020thesis}. The deposited energy for each simulation run is $\SI{0.12}{eV}$, deposited uniformly in a circular area with radius \SI{10}{\nano\meter}, which represents the photon energy minus athermal phonon losses during down-conversion.





\end{appendices}


\bibliography{sn-bibliography}

@article{Sidorova2017,
  author  = {Sidorova, Mariia and Semenov, Alexej and Hubers, Heinz-Wilhelm and Charaev, Ilya and Kuzmin, Artem and Doerner, Steffen and Siegel, Michael},
  title   = {Physical mechanisms of timing jitter in photon detection by current
  carrying superconducting nanowires},
  journal = {Phys. Rev. B 96, 184504},
  year    = {2017},
  doi     = {10.1103/PhysRevB.96.184504}
}

@article{Haneishi2025,
  author = {Haneishi, Tomohiro and Yabuno, Masahiro and Kutsuma, Hiroki and Miki, Shigehito and Yamashita, Taro},
  title = {Evaluation of NbTiN Superconducting Strip Photon Detectors With 0.1-30 $\mu$m Strip Widths},
  journal = {IEEE Transactions on Applied Superconductivity},
  volume = {35},
  number = {5},
  pages = {2200105},
  year = {2025},
  doi = {10.1109/TASC.2024.3521901}
}

@article{Hughes2026,
    author = {Hughes, Emi Cora Valmai and Upadhya, Avinash and Dholakia, Kishan},
    title = {Superconducting nanowire single-photon detectors for enhanced biomedical imaging},
    journal = {Journal of Biomedical Optics},
    volume = {31},
    number = {11},
    pages = {113502},
    year = {2026},
    publisher = {SPIE},
    doi = {10.1117/1.JBO.31.11.113502}
}

@article{Grunenfelder2023,
    author = {Gr\"{u}nenfelder, Fadri and Boaron, Alberto and Resta, Giovanni V. and Perrenoud, Matthieu and Rusca, Davide and Barreiro, Claudio and Houlmann, Rapha\"{e}l and Sax, Rebecka and Stasi, Lorenzo and El-Khoury, Sylvain and Zbinden, Hugo},
    title = {Fast single-photon detectors and real-time key distillation enable high secret-key-rate quantum key distribution systems},
    journal = {Nature Photonics},
    volume = {17},
    number = {5},
    pages = {422--426},
    year = {2023},
    publisher = {Nature Publishing Group},
    doi = {10.1038/s41566-023-01168-2}
}

@article{Chen2021,
    author = {Chen, J.-P. and Zhang, C. and Liu, Y. and Jiang, C. and Zhang, W.-J. and Han, Z.-Y. and Ma, S.-Z. and Hu, X.-L. and Li, Y.-H. and Liu, H. and Zhou, F. and Jiang, H.-F. and Chen, T.-Y. and Li, H. and You, L.-X. and Wang, Z. and Wang, X.-B. and Zhang, Q. and Pan, J.-W.},
    title = {Twin-field quantum key distribution over a 511 km optical fibre linking two distant metropolitan areas},
    journal = {Nature Photonics},
    volume = {15},
    number = {8},
    pages = {570--575},
    year = {2021},
    publisher = {Nature Publishing Group},
    doi = {10.1038/s41566-021-00828-5}
}

@article{Cheng2023,
    author = {Cheng, R. and Zhou, Y. and Wang, S. and Shen, M. and Taher, T. and Tang, H. X.},
    title = {A 100-pixel photon-number-resolving detector unveiling photon statistics},
    journal = {Nature Photonics},
    volume = {17},
    number = {1},
    pages = {112--119},
    year = {2023},
    publisher = {Nature Publishing Group},
    doi = {10.1038/s41566-022-01119-3}
}

@article{Zhong2020,
    author = {Zhong, H.-S. and Wang, H. and Deng, Y.-H. and Chen, M.-C. and Peng, L.-C. and Luo, Y.-H. and Qin, J. and Wu, D. and Ding, X. and Hu, Y. and Hu, P. and Yang, X.-Y. and Zhang, W.-J. and Li, H. and Li, Y. and Jiang, X. and Gan, L. and Yang, G. and You, L. and Wang, Z. and Li, L. and Liu, N.-L. and Lu, C.-Y. and Pan, J.-W.},
    title = {Quantum computational advantage using photons},
    journal = {Science},
    volume = {370},
    number = {6523},
    pages = {1460--1463},
    year = {2020},
    publisher = {American Association for the Advancement of Science},
    doi = {10.1126/science.abe8770}
}

@article{Skocpol1976,
    author = {Skocpol, W. J.},
    title = {Critical currents of superconducting microbridges},
    journal = {Physical Review B},
    volume = {14},
    number = {3},
    pages = {1045--1051},
    year = {1976},
    publisher = {American Physical Society},
    doi = {10.1103/PhysRevB.14.1045}
}

@article{Calandri2016,
    author = {Calandri, N. and Zhao, Q. Y. and Zhu, D. and Dane, A. and Berggren, K. K.},
    title = {Superconducting nanowire detector jitter limited by detector geometry},
    journal = {Applied Physics Letters},
    volume = {109},
    number = {15},
    pages = {152601},
    year = {2016},
    publisher = {AIP Publishing},
    doi = {10.1063/1.4963158}
}

@article{Bulaevskii2011,
  author  = {Bulaevskii, L. N. and Graf, M. J. and Batista, C. D. and Kogan, V. G.},
  title   = {Vortex-induced dissipation in narrow current-biased thin-film superconducting strips},
  journal = {Physical Review B},
  year    = {2011},
  volume  = {83},
  number  = {14},
  pages   = {144526},
  doi     = {10.1103/PhysRevB.83.144526}
}

@article{Clem2011,
  author  = {Clem, John R. and Berggren, Karl K.},
  title   = {Geometry-dependent critical currents in superconducting nanocircuits},
  journal = {Physical Review B},
  year    = {2011},
  volume  = {84},
  number  = {17},
  pages   = {174510},
  doi     = {10.1103/PhysRevB.84.174510}
}

@article{Renema2015,
  author  = {Renema, Jelmer J. and Wang, Qiang and Gaudio, Roberto and Komen, I. and op 't Hoog, K. and Sahin, D. and Schilling, A. and van Exter, Martin P. and Fiore, Andrea and Engel, Andreas and de Dood, Michiel J. A.},
  title   = {Position-Dependent Local Detection Efficiency in a Nanowire Superconducting Single-Photon Detector},
  journal = {Nano Letters},
  year    = {2015},
  volume  = {15},
  number  = {7},
  pages   = {4541--4545},
  doi     = {10.1021/acs.nanolett.5b01103}
}

@article{Zhao2017,
  author  = {Zhao, Qing-Yuan and Zhu, Di and Calandri, Niccolò and Dane, Andrew E. and McCaughan, Adam N. and Bellei, Francesco and Wang, Hao-Zhu and Santavicca, Daniel F. and Berggren, Karl K.},
  title   = {Single-photon imager based on a superconducting nanowire delay line},
  journal = {Nature Photonics},
  year    = {2017},
  volume  = {11},
  pages   = {247--251},
  doi     = {10.1038/nphoton.2017.35}
}

@article{Cahall2017,
    author = {Cahall, C. and Nicolich, K. L. and Islam, N. T. and Lafyatis, G. P. and Miller, A. J. and Gauthier, D. J. and Kim, J.},
    title = {Multi-photon detection using a conventional superconducting nanowire single-photon detector},
    journal = {Optica},
    volume = {4},
    number = {12},
    pages = {1534--1535},
    year = {2017},
    publisher = {Optica Publishing Group},
    doi = {10.1364/OPTICA.4.001534}
}

@article{Nicolich2019,
    author = {Nicolich, K. L. and Cahall, C. and Islam, N. T. and Lafyatis, G. P. and Kim, J. and Miller, A. J. and Gauthier, D. J.},
    title = {Universal photon-number-resolving detector based on a single superconducting nanowire},
    journal = {Nature Photonics},
    volume = {13},
    number = {12},
    pages = {845--849},
    year = {2019},
    publisher = {Nature Publishing Group},
    doi = {10.1038/s41566-019-0526-8}
}

@article{McCaughan2022_TCI,
    author = {McCaughan, Adam N. and Zhai, Y. and Korzh, B. and Allmaras, J. P. and Oripov, B. G. and Shaw, M. D. and Nam, S. W.},
    title = {The thermally coupled imager: A scalable readout architecture for superconducting nanowire single photon detectors},
    journal = {Applied Physics Letters},
    volume = {121},
    number = {10},
    pages = {102602},
    year = {2022},
    doi = {10.1063/5.0102154},
    publisher = {AIP Publishing}
}

@article{Yabuno2023,
    author = {Yabuno, Masahiro and China, Fumihiro and Miki, Shigehito and Yamashita, Taro and Terai, Hirotaka},
    title = {Superconducting wide strip photon detector with high critical current},
    journal = {Optica Quantum},
    volume = {1},
    number = {1},
    pages = {26--34},
    year = {2023},
    publisher = {Optica Publishing Group},
    doi = {10.1364/OPTICAQ.497555}
}

@article{allmaras2019intrinsic,
    author = {Allmaras, Jason P. and Kozorezov, Alexander G. and Korzh, Boris A. and Beyer, Andrew D. and Shaw, Matthew D.},
    title = {Intrinsic timing jitter and latency in superconducting nanowire single-photon detectors},
    journal = {Nano Letters},
    volume = {19},
    number = {4},
    pages = {2685--2691},
    year = {2019},
    publisher = {ACS Publications},
    doi = {10.1021/acs.nanolett.8b04746}
}

@article{Mueller2021,
    author = {Mueller, Andrew S. and Korzh, Boris and Runyan, Marcus and Wollman, Emma E. and Beyer, Andrew D. and Allmaras, Jason P. and Velasco, Angel E. and Craiciu, Ioana and Bumble, Bruce and Briggs, Ryan M. and Narvaez, Lautaro and Pe{\~{n}}a, Cristi{\'{a}}n and Spiropulu, Maria and Shaw, Matthew D.},
    title = {Free-space coupled superconducting nanowire single-photon detector with low dark counts},
    journal = {Optica},
    volume = {8},
    number = {12},
    pages = {1586--1587},
    year = {2021},
    doi = {10.1364/OPTICA.444108}
}

@article{Korzh2020,
    author = {Korzh, Boris and Zhao, Qing-Yuan and Allmaras, Jason P. and Frasca, Simone and Autry, Travis M. and Bersin, Eric A. and Beyer, Andrew D. and Briggs, Ryan M. and Bumble, Bruce and Colangelo, Marco and Crouch, G. M. and Dane, A. E. and Gerrits, T. and Lita, A. E. and Marsili, F. and Moody, G. and Pe{\~{n}}a, C. and Ramirez, E. and Rezac, J. D. and Sinclair, N. and Stevens, M. J. and Velasco, A. E. and Verma, V. B. and Wollman, E. E. and Xie, S. and Zhu, D. and Hale, P. D. and Spiropulu, M. and Silverman, K. L. and Mirin, R. P. and Nam, S. W. and Kozorezov, A. G. and Shaw, M. D. and Berggren, K. K.},
    title = {Demonstration of sub-3 ps temporal resolution with a superconducting nanowire single-photon detector},
    journal = {Nature Photonics},
    volume = {14},
    number = {4},
    pages = {250--255},
    year = {2020},
    publisher = {Nature Publishing Group},
    doi = {10.1038/s41566-020-0589-x}
}

@article{Chang2021,
    author = {Chang, Jin and Los, Johannes W. N. and Tenorio-Pearl, J. O. and Noordzij, N. and Gourgues, Ronan B. M. and Guardiani, Antonio and Zichi, Julien Romain and Pereira, Silvania F. and Urbach, H. P. and Zwiller, Val and Dorenbos, Sander N. and Zadeh, Iman Esmaeil},
    title = {Detecting telecom single photons with 99.5-2.07+ 0.5\% system detection efficiency and high time resolution},
    journal = {APL Photonics},
    volume = {6},
    number = {3},
    pages = {036114},
    year = {2021},
    publisher = {AIP Publishing},
    doi = {10.1063/5.0039772}
}

@article{Reddy2020,
    author = {Reddy, Dileep V. and Nerem, Robert R. and Nam, Sae Woo and Mirin, Richard P. and Verma, Varun B.},
    title = {Superconducting nanowire single-photon detectors with 98\% system detection efficiency at 1550 nm},
    journal = {Optica},
    volume = {7},
    number = {12},
    pages = {1649--1653},
    year = {2020},
    publisher = {Optica Publishing Group},
    doi = {10.1364/OPTICA.400751}
}

@article{Chiles2022,
    author = {Chiles, Jeff and Charaev, Ilya and Lasenby, Robert and Baryakhtar, Masha and Huang, Junwu and Roshko, Alexana and Burton, George and Colangelo, Marco and Van Tilburg, Ken and Arvanitaki, Asimina and Nam, Sae Woo and Berggren, Karl K.},
    title = {New constraints on dark photon dark matter with superconducting nanowire detectors in an optical haloscope},
    journal = {Physical Review Letters},
    volume = {128},
    number = {23},
    pages = {231802},
    year = {2022},
    publisher = {American Physical Society},
    doi = {10.1103/PhysRevLett.128.231802}
}

@article{Korneeva2018,
    author = {Korneeva, Yuliya P. and Florya, Ilya N. and Vachtomin, Yury B. and Smirnov, Konstantin V. and Divochiy, Aleksander V. and Morozov, Pavel V. and Seleznev, Vyacheslav A. and Gol'tsman, Gregory N. and Klushin, Alexander M.},
    title = {Optical single-photon detection in micrometer-wide {NbN} strips},
    journal = {Physical Review Applied},
    volume = {9},
    number = {6},
    pages = {064037},
    year = {2018},
    publisher = {American Physical Society},
    doi = {10.1103/PhysRevApplied.9.064037}
}

@misc{Gurevich2026,
    author = {Gurevich, Alex},
    title = {Tuning current flow in superconducting thin film strips by control wires. Applications to single photon detectors and diodes.},
    year = {2026},
    eprint = {2602.02984},
    archivePrefix = {arXiv},
    primaryClass = {cond-mat.supr-con},
    url = {https://arxiv.org/abs/2602.02984}
}

@article{Parzuchowski2026,
    author = {Parzuchowski, Kristen M. and Mueller, Eli and Oripov, Bakhrom G. and Hampel, Benedikt and Chowdhury, Ravin A. and Patel, Sahil R. and Kuznesof, Daniel and Batson, Emma K. and Morgenstern, Ryan and Hadfield, Robert H. and Verma, Varun B. and Shaw, Matthew D. and Allmaras, Jason P. and Stevens, Martin J. and Gurevich, Alex and McCaughan, Adam N.},
    title = {Reaching the intrinsic performance limits of superconducting nanowire single-photon detectors up to 0.1 mm wide},
    journal = {Optica},
    volume = {13},
    number = {8},
    pages = {1649--1657},
    year = {2026},
    publisher = {Optica Publishing Group}
}

@article{kuznesof2025midinfrared,
    author = {Kuznesof, Daniel and Amer, Noureldin and Taylor, Gregor G. and Patel, Sahil R. and Bumble, Bruce and Craiciu, Ioana and Choudhary, Nidhi and Choi, Beomgyu and Morozov, Dmitry V. and Korzh, Boris and Shaw, Matthew D. and Wollman, Emma E. and Chong, Yonuk and Hadfield, Robert H.},
    title = {Mid-infrared high-resolution photon-counting {LiDAR}},
    journal = {Optics Express},
    volume = {33},
    number = {22},
    pages = {45684--45694},
    year = {2025},
    publisher = {Optica Publishing Group},
    doi = {10.1364/OE.576861}
}

@article{Zadeh2020,
    author = {Zadeh, Iman Esmaeil and Los, Johannes W. N. and Gourgues, Ronan B. M. and Chang, Jin and Elshaari, Ali W. and Zichi, Julien Romain and Van Staaden, Yuri J. and Swens, Jeroen P. E. and Kalhor, Nima and Guardiani, Antonio and Meng, Yun and Zou, Kai and Dobrovolskiy, Sergiy and Fognini, Andreas W. and Schaart, Dennis R. and Dalacu, Dan and Poole, Philip J. and Reimer, Michael E. and Hu, Xiaolong and Pereira, Silvania F. and Zwiller, Val and Dorenbos, Sander N.},
    title = {Efficient Single-Photon Detection with 7.7 ps Time Resolution for Photon-Correlation Measurements},
    journal = {ACS Photonics},
    volume = {7},
    number = {7},
    pages = {1780--1787},
    year = {2020},
    publisher = {American Chemical Society},
    doi = {10.1021/acsphotonics.0c00433}
}

@article{vodolazov2017single,
    author = {Vodolazov, D. Yu.},
    title = {Single-Photon Detection by a Dirty Current-Carrying Superconducting Strip Based on the Kinetic-Equation Approach},
    journal = {Physical Review Applied},
    volume = {7},
    number = {3},
    pages = {034014},
    year = {2017},
    month = {Mar},
    publisher = {American Physical Society},
    doi = {10.1103/PhysRevApplied.7.034014}
}

@article{Wang2022,
    author = {Wang, Feifei and Ren, Fuqiang and Ma, Zhuoran and Qu, Liangqiong and Gourgues, Ronan and Xu, Chun and Baghdasaryan, Ani and Li, Jiachen and Zadeh, Iman Esmaeil and Los, Johannes and Fognini, Andreas and Qin-Dregely, Jessie and Dai, Hongjie},
    title = {In vivo non-invasive confocal fluorescence imaging beyond 1,700 nm using superconducting nanowire single-photon detectors},
    journal = {Nature Nanotechnology},
    volume = {17},
    pages = {653--660},
    year = {2022},
    month = {Jun},
    publisher = {Nature Publishing Group},
    doi = {10.1038/s41565-022-01130-3}
}

@article{Hochberg2022,
    author = {Hochberg, Yonit and Lehmann, Benjamin V. and Charaev, Ilya and Chiles, Jeff and Colangelo, Marco and Nam, Sae Woo and Berggren, Karl K.},
    title = {New constraints on dark matter from superconducting nanowires},
    journal = {Physical Review D},
    volume = {106},
    number = {11},
    pages = {112005},
    year = {2022},
    month = {Dec},
    publisher = {American Physical Society},
    doi = {10.1103/PhysRevD.106.112005}
}

@article{You2020,
    author = {You, Lixing},
    title = {Superconducting nanowire single-photon detectors for quantum information},
    journal = {Nanophotonics},
    volume = {9},
    number = {9},
    pages = {2673--2692},
    year = {2020},
    publisher = {De Gruyter},
    doi = {10.1515/nanoph-2020-0186}
}

@article{Hao2024,
    author = {Hao, Hao and Zhao, Qing-Yuan and Huang, Yang-Hui and Deng, Jie and Yang, Fan and Ru, Sai-Ying and Liu, Zhen and Wan, Chao and Liu, Hao and Li, Zhi-Jian and Wang, Hua-Bing and Tu, Xue-Cou and Zhang, La-Bao and Jia, Xiao-Qing and Wu, Xing-Long and Chen, Jian and Kang, Lin and Wu, Pei-Heng},
    title = {A compact multi-pixel superconducting nanowire single-photon detector array supporting gigabit space-to-ground communications},
    journal = {Light: Science \& Applications},
    volume = {13},
    number = {1},
    pages = {25},
    year = {2024},
    month = {Dec},
    publisher = {Nature Publishing Group},
    doi = {10.1038/s41377-023-01374-1}
}

@article{Wollman2024,
    author = {Wollman, Emma E. and Allmaras, Jason P. and Beyer, Andrew D. and Korzh, Boris and Runyan, Marc C. and Narv{\'{a}}ez, Lautaro and Farr, William H. and Marsili, Francesco and Briggs, Ryan M. and Miles, Gregory J. and Shaw, Matthew D.},
    title = {{SNSPD}-based detector system for {NASA's} Deep Space Optical Communications project},
    journal = {Optics Express},
    volume = {32},
    number = {27},
    pages = {48185--48198},
    year = {2024},
    month = {Dec},
    publisher = {Optica Publishing Group},
    doi = {10.1364/OE.541425}
}

@inproceedings{Allmaras2017,
    author = {Allmaras, J. P. and Beyer, A. D. and Briggs, R. M. and Marsili, F. and Shaw, M. D. and Resta, G. V. and Stern, J. A. and Verma, V. B. and Mirin, R. P. and Nam, S. W. and Farr, W. H.},
    title = {Large-Area 64-pixel Array of {WSi} Superconducting Nanowire Single Photon Detectors},
    booktitle = {Conference on Lasers and Electro-Optics},
    pages = {JTh3E.7},
    year = {2017},
    publisher = {Optica Publishing Group},
    address = {San Jose, California},
    doi = {10.1364/CLEO_AT.2017.JTh3E.7}
}

@article{Hadfield2023,
    author = {Hadfield, Robert H. and Leach, Jonathan and Fleming, Fiona and Paul, Douglas J. and Tan, Chee Hing and Ng, Jo Shien and Henderson, Robert K. and Buller, Gerald S.},
    title = {Single-photon detection for long-range imaging and sensing},
    journal = {Optica},
    volume = {10},
    number = {9},
    pages = {1124--1141},
    year = {2023},
    month = {Sep},
    publisher = {Optica Publishing Group},
    doi = {10.1364/OPTICA.488853}
}

@article{McCarthy2025,
    author = {McCarthy, Aongus and Taylor, Gregor G. and Garcia-Armenta, Jorge and Korzh, Boris and Morozov, Dmitry V. and Beyer, Andrew D. and Briggs, Ryan M. and Allmaras, Jason P. and Bumble, Bruce and Colangelo, Marco and Zhu, Di and Berggren, Karl K. and Shaw, Matthew D. and Hadfield, Robert H. and Buller, Gerald S.},
    title = {High-resolution long-distance depth imaging {LiDAR} with ultra-low timing jitter superconducting nanowire single-photon detectors},
    journal = {Optica},
    volume = {12},
    number = {2},
    pages = {168--177},
    year = {2025},
    month = {Feb},
    publisher = {Optica Publishing Group},
    doi = {10.1364/OPTICA.544877}
}

@article{Luskin2023,
    author = {Luskin, Jamie S. and Schmidt, Ekkehart and Korzh, Boris and Beyer, Andrew D. and Bumble, Bruce and Allmaras, Jason P. and Walter, Alexander B. and Wollman, Emma E. and Narv{\'{a}}ez, Lautaro and Verma, Varun B. and Nam, Sae Woo and Charaev, Ilya and Colangelo, Marco and Berggren, Karl K. and Pe{\~{n}}a, Cristi{\'{a}}n and Spiropulu, Maria and Garc{\'{i}}a-Sciveres, Maurice and Derenzo, Stephen and Shaw, Matthew D.},
    title = {Large active-area superconducting microwire detector array with single-photon sensitivity in the near-infrared},
    journal = {Applied Physics Letters},
    volume = {122},
    number = {24},
    pages = {243506},
    year = {2023},
    month = {Jun},
    publisher = {AIP Publishing},
    doi = {10.1063/5.0150282}
}

@article{Chiles2020,
    author = {Chiles, J. and Buckley, S. and Lita, A. and Verma, V. and Shainline, J. and Mirin, R. and Nam, S. and Allmaras, J. and Korzh, B. and Wollman, E. and Shaw, M.},
    title = {Superconducting microwire detectors with single-photon sensitivity in the near-infrared},
    journal = {Applied Physics Letters},
    volume = {116},
    number = {24},
    pages = {242602},
    year = {2020},
    publisher = {AIP Publishing},
    doi = {10.1063/5.0006221}
}

@phdthesis{allmaras2020thesis,
    author = {Allmaras, Jason Paul},
    title = {Modeling and development of superconducting nanowire single-photon detectors},
    school = {California Institute of Technology},
    year = {2020},
    doi = {10.7907/vntw-f597}
}

@article{Vodolazov2019,
    author = {Vodolazov, D. Yu.},
    title = {Minimal Timing Jitter in Superconducting Nanowire Single-Photon Detectors},
    journal = {Physical Review Applied},
    volume = {11},
    number = {1},
    pages = {014016},
    year = {2019},
    month = {Jan},
    publisher = {American Physical Society},
    doi = {10.1103/PhysRevApplied.11.014016}
}

@article{Vodolazov2020_uWire,
  title = {Timing Jitter in {NbN} Superconducting Microstrip Single-Photon Detector},
  author = {Vodolazov, D.Yu. and Manova, N.N. and Korneeva, Yu.P. and Korneev, A.A.},
  journal = {Phys. Rev. Appl.},
  volume = {14},
  issue = {4},
  pages = {044041},
  numpages = {8},
  year = {2020},
  month = {Oct},
  publisher = {American Physical Society},
  doi = {10.1103/PhysRevApplied.14.044041},
  url = {https://link.aps.org/doi/10.1103/PhysRevApplied.14.044041}
}

\end{document}